\documentclass{article}

\renewcommand{\baselinestretch}{1.05}

\usepackage[utf8]{inputenc}
\usepackage[T1]{fontenc}

\usepackage{graphicx,caption}
\usepackage{ragged2e}
\usepackage{amsmath,amssymb,amsfonts,bbm}
\usepackage{times,txfonts}
\usepackage[mathscr]{euscript}

\usepackage[scaled=1]{helvet}
\usepackage[helvet]{sfmath}

\usepackage[left=4cm,top=2.5cm,right=4cm,bottom=2.5cm]{geometry}

\usepackage{hyperref}

\usepackage{braket}

\begin{document}

\title{\Large \bfseries An introduction to quantum discord and non-classical correlations beyond entanglement}
\author{{\large Gerardo Adesso, Marco Cianciaruso, and Thomas R. Bromley} \\[2pt] 
{\small \em Centre for the Mathematics and Theoretical Physics of Quantum Non-Equilibrium Systems,} \\[-2pt]
{\small \em School of Mathematical Sciences, The University of Nottingham,} \\[-2pt]
{\small \em University Park, Nottingham NG7 2RD, United Kingdom}}
\date{}

\maketitle
\section{\!\!\!Introduction}

What is quantum? As researchers of quantum physics, we are constantly bombarded with attributes like ``non-classical'' and ``super-classical''. We strive to track down the elusive quantum-classical boundary, to understand what makes quantum mechanics so powerful yet counter-intuitive. But to do this, we must first have a firm understanding of the classical world and the laws that classical mechanics imposes.

There are in fact many ways to think about classicality. One facet of the classical world is that any system is always in a fixed and predetermined state. Take for example a bit: it can be either 0 or 1. How does this compare with what is predicted from the rulebook of quantum mechanics? Here we can have systems existing in a superposition of both 0 and 1, called quantum bits or qubits. This form of non-classicality is what is known as quantum coherence~\cite{streltsov2016review}.

It is also interesting to consider systems of spatially separated parties and the correlations between them. We can try to identify the states that are describable by classical mechanics and the states that are not. You are probably now thinking that this sounds a lot like entanglement~\cite{horodecki2009quantum}, and that the classically correlated states are just separable states. However, things are not so simple: it turns out that even separable mixed states can exhibit some quantumness in their correlations!

In this manuscript we will explore these manifestations of quantum correlations \emph{beyond} entanglement \cite{modi2012classical,streltsov2015quantum,adesso2016measures}. We begin by introducing and motivating the classically correlated states and then showing how to quantify the quantum correlations using an entropic approach, arriving at a well known measure called the quantum discord~\cite{ollivier2001quantum,henderson2001classical}. Quantum correlations and discord are then operationally linked with the task of local broadcasting~\cite{piani2008no}. 
We conclude by providing some alternative perspectives on quantum correlations and how to measure them.

Finally, before proceeding it is important to note that there are many layers of quantumness in composite systems. As well as entanglement and discord-type quantum correlations, one can identify e.g.~steering and Bell non-locality.  For pure composite states, all of these signatures of quantumness become equivalent, yet for mixed states they are different, showing a strict hierarchy. Each form of quantumness is of independent interest, but here we focus on the most general form of quantum correlations, leaving the interested reader to consult Ref.~\cite{horodecki2009quantum} for more information on entanglement and Refs.~\cite{cavalcanti2016quantum,brunner2014bell} for steering and non-locality.

\section{\!\!\!Quantumness  versus classicality (of correlations)}


Generally, {\em quantumness} can represent any of the counter-intuitive phenomena that we encounter when investigating microscopic systems such as atoms, electrons, photons, and many others. In particular, the quantumness of \emph{correlations} manifests itself when two such microscopic systems interact with each other, and stands as one of the weirdest of all quantum features. In order to really appreciate any sort of quantumness, we first need to thoroughly understand how the classical world works, i.e., we first need to agree on what exactly ``intuitive'' means, and only afterwards benchmark quantummness against such a standard. This is the purpose of this section.


Let us set the stage for our comparison of the classical and the quantum. From a minimalistic point of view, both classical and quantum systems can be described by resorting to the following four ingredients: the set of states, the set of observables, a real number associated with any pairing of a state and observable, which is the predicted result of a measurement of the given observable when the system is in the given state, and a family of mappings describing the dynamics of the system. However, in the following we will focus only on the first three ingredients; we will also specialize to discrete variable systems for the sake of simplicity.

The state of a discrete variable {\it classical} system, whose phase space $\mathscr{M}$ is formed by $d$ points that we label by ${\{i\}}_{i=1}^d$, can be described by a probability distribution $\textbf{p}={\{p_i\}}_{i=1}^d$ defined on $\mathscr{M}$, i.e., any set of $d$ numbers that are non-negative, $p_i\geq 0$, and normalized, $\sum_{i=1}^d p_i = 1$. An observable of such a system is instead any real function $\textbf{f}={\{f_i\}}_{i=1}^d$ on $\mathscr{M}$, i.e., $f_i^*=f_i$, while what we actually observe by measuring the observable $\textbf{f}$ when the system is in the state $\textbf{p}$ is the corresponding expectation value, i.e., ${\langle \textbf{f} \rangle}_\textbf{p} = \textbf{p}\cdot\textbf{f} = \sum_{i=1}^d p_i f_i$.

We say that a classical system is in a pure state when we have the best possible knowledge about it, i.e., we know with certainty what point of the phase space is occupied by the system. In fact, pure states of classical systems are nothing but Kronecker deltas ${\{\delta_{ik}\}}_{i=1}^d$, with $k$ being the point in the phase space occupied by the system, i.e., $\delta_{ik}=1$ if $i=k$ while $\delta_{ik}=0$ if $i\neq k$. Moreover, when a classical system is in a pure state ${\{\delta_{ik}\}}_{i=1}^d$ we can predict with certainty that the result of the measurement of an arbitrary observable $\textbf{f}$ is the value $f_k$, where $k$ is the point of the phase space occupied by the system. Interestingly, every state of a classical system that is not pure can be obtained in a unique way as a convex combination of pure states and it is thus called a mixed state.

Our ignorance about the state $\textbf{p}$ of a classical system can be quantified by resorting to its Shannon entropy,
\begin{equation}
\mathcal{S}(\textbf{p})=-\sum_{i=1}^d p_i \log p_i,
\end{equation}
which is indeed zero for pure states and reaches its maximum for the so-called maximally mixed state, which is such that $p_i=1/d$ for any $i$ and thus entails that we have the least possible knowledge about which one of the points of the phase space is actually occupied by the system, being them all equally probable.


When considering two discrete variable classical subsystems $A$ and $B$, with phase spaces given by $\mathscr{M}^A={\{i\}}_{i=1}^{d_A}$ and $\mathscr{M}^B={\{j\}}_{j=1}^{d_B}$, respectively, it happens that the phase space $\mathscr{M}^{AB}$ corresponding to the composite system $AB$ is the Cartesian product of the ones corresponding to the two subsystems, i.e., $\mathscr{M}^{AB}=\mathscr{M}^A \times \mathscr{M}^B$, whose points are given by the $d_Ad_B$ ordered pairs ${\{(i,j)\}}_{i,j=1}^{d_A,d_B}$. The state of a bipartite classical system can be thus described by a joint probability distribution $\textbf{p}^{AB}={\lbrace p_{ij}^{AB}\rbrace}_{i,j=1}^{d_A,d_B}$ defined on $\mathscr{M}^{AB}$, while the states of the subsystems $A$ and $B$ can be characterized by the corresponding marginal probability distributions, i.e., $\textbf{p}^{A}={\lbrace p_{i}^{A} = \sum_{j=1}^{d_B} p_{ij}^{AB} \rbrace}_{i=1}^{d_A}$ and $\textbf{p}^{B}={\{p_{j}^{B} = \sum_{i=1}^{d_A} p_{ij}^{AB}\}}_{j=1}^{d_B}$, respectively.

In particular, pure states of bipartite classical systems are given by products of Kronecker deltas, ${\{\delta_{ik}\delta_{jl}\}}_{i,j=1}^{d_A,d_B}$, where $(k,l)$ is the point of the phase space occupied with certainty by the bipartite system, i.e., $\delta_{ik}\delta_{jl}=1$ if $(i,j)=(k,l)$ while $\delta_{ik}\delta_{jl}=0$ if $(i,j)\neq(k,l)$. Again, every state of a bipartite classical state that is not pure can be written in a unique way as a convex combination of pure bipartite states, i.e., as a classical mixture of products of Kronecker deltas. Furthermore, quite interestingly, when a bipartite classical system is in a pure state, then also the subsystems are necessarily in a pure state, indeed one can easily see that the marginal distributions of ${\{\delta_{ik}\delta_{jl}\}}_{i,j=1}^{d_A,d_B}$ are $\{\delta_{ik}\}_{i=1}^{d_A}$ and $\{\delta_{jk}\}_{j=1}^{d_B}$. In other words, within the classical world, if we have the best possible knowledge of the state of a composite system, then we necessarily have the best possible knowledge of the states of both its subsystems.

On the other hand, the state of a discrete variable {\it quantum} system, whose Hilbert space $\mathscr{H}$ has a finite dimension $d$, can be described by a density operator $\rho$ acting on $\mathscr{H}$, i.e., any linear operator on $\mathscr{H}$ that is positive semi-definite, $\rho\geq 0$, and normalized, $\mbox{Tr}(\rho)=1$. An observable of such a system is instead any Hermitian operator $O$ on $\mathscr{H}$, i.e., $O^\dagger = O$, while what we actually observe by measuring the observable $O$ when the system is in the state $\rho$ is the corresponding expectation value, i.e., $\langle O \rangle_\rho = \mbox{Tr}(\rho O)$.

Again, we say that a quantum system is in a pure state when we have the best possible knowledge about it, i.e., we know with certainty what normalized vector of the Hilbert space is occupied by the system. Pure states of quantum systems are thus described by projectors $|\psi\rangle\langle\psi|$ onto normalized vectors $|\psi\rangle$ of $\mathscr{H}$. Moreover, when a quantum system is in a pure state $|\psi\rangle$, we can predict with certainty the result of the measurement of any observable $O$ having $|\psi\rangle$ between its eigenvectors, without perturbing the state of the system whatsoever. However, contrary to what happens in the classical world, this is no longer the case when we measure any other kind of observable, whose eigenvectors are different from $|\psi\rangle$. More precisely, if we measure a generic observable $O$ with eigenvectors $\{\Pi_i\}$ when the quantum system is in the state $\rho$, it happens that the state of the system can collapse onto any of the eigenstates $\Pi_i$ of $O$ with probability $p_i=\mbox{Tr}(\rho \Pi_i)$. This is not due to our ignorance about the state of the system, but rather to an intrinsic indeterminism manifested by nature at the microscopic level, a fact which stands
as one of the most striking features of quantumness. This phenomenon is mathematically taken into account by the fact that in the quantum setting we have that states and observables are no longer commuting real functions but rather possibly non-commuting Hermitian operators.

Yet there is another striking quantum feature that manifests itself in single quantum systems, as we have already alluded to: the celebrated quantum superposition, or coherence. It arises from the fact that in the quantum setting we are not only allowed to consider classical mixtures of pure states, i.e., $\rho=\sum_i p_i |\psi_i\rangle\langle\psi_i|$, also called simply mixed states, but rather we can also construct coherent superpositions of pure states that give rise to other pure states, i.e., $|\psi\rangle = \sum_i c_i |\psi_i\rangle$. However, particular mention has to be given to superpositions and mixtures of elements of an orthonormal basis ${\{|i\rangle\}}_{i=1}^d$ of $\mathscr{H}$. Indeed, one can easily appreciate that, due to the perfect distinguishability of orthogonal states, any quantum state of the form $\sum_{i=1}^d p_i |i\rangle\langle i|$ can be simulated by the classical state ${\{p_i\}}_{i=1}^d$. Therefore, such states represent a stereotype of classicality within the quantum world and are called incoherent states.

Our classical ignorance about the state $\rho$ of a quantum system can be quantified by resorting to its von Neumann entropy,
\begin{equation}
\mathcal{S}(\rho)=-\mbox{Tr}(\rho\log\rho),
\end{equation}
which is indeed zero for pure states and reaches its maximum for the maximally mixed state, $\mathbb{I}/d$, with $\mathbb{I}$ being the identity on $\mathscr{H}$.

When considering two discrete variable quantum systems $A$ and $B$, with Hilbert spaces given by $\mathscr{H}^A$ and $\mathscr{H}^B$, respectively, it happens that the Hilbert space $\mathscr{H}^{AB}$ corresponding to the composite system $AB$ is the tensor product of the ones corresponding to the two subsystems, i.e., $\mathscr{H}^{AB}=\mathscr{H}^A\otimes\mathscr{H}^B$, which is a $(d_Ad_B)$-dimensional Hilbert space whose vectors are spanned by the orthonormal product basis ${\{|i^A\rangle\otimes |j^B\rangle\}}_{i,j=1}^{d_A,d_B}$, with ${\{|i^A\rangle\}}_{i=1}^{d_A}$ and ${\{|j^B\rangle\}}_{i=1}^{d_B}$ being orthonormal bases of $\mathscr{H}^A$ and $\mathscr{H}^B$, respectively. The state of a bipartite quantum system can be thus described by a density operator $\rho^{AB}$ acting on $\mathscr{H}^{AB}$, while the states of the subsystems $A$ and $B$ can be characterized by the corresponding marginal density operators, i.e., $\rho^A=\mbox{Tr}_B(\rho^{AB})$ and $\rho^B=\mbox{Tr}_A(\rho^{AB})$, respectively, where $\mbox{Tr}_X$ is the partial trace over the Hilbert space of subsystem $X$.

In particular, pure states of bipartite quantum systems are given by projectors onto normalized vectors of $\mathscr{H}^{AB}$. Here comes one of the most amazing features of quantum mechanics, which is attributed to quantum correlations. Due to both the superposition principle and the tensorial structure of the  Hilbert space of the composite system, it happens that a pure bipartite quantum state is not necessarily factorizable in the tensor product of two pure states of the subsystems, i.e., $|\psi^{AB}\rangle$ cannot be written in general in the form $|\phi^A\rangle\otimes|\varphi^B\rangle$, with $|\phi^A\rangle\in\mathscr{H}^A$ and $|\varphi^B\rangle\in\mathscr{H}^B$. An immediate consequence of the non-factorizability of a pure bipartite state $|\psi^{AB}\rangle$ is the fact that the corresponding subsystems' states are necessarily non-pure. In other words, within the quantum world, the best possible knowledge of the state of a composite system does not imply the best possible knowledge of the states of the two subsystems. This is in stark contrast with what happens in the classical world and, as Schr\"{o}dinger said, stands as ``not \textit{one} but rather \textit{the} characteristic trait of quantum mechanics, the one that enforces its entire departure from classical line of thought'' \cite{schrodinger1935discussion}. This phenomenon was baptized {\it entanglement} by Schr\"{o}dinger, but it is nowadays more broadly known as quantum correlations for pure states. Overall, for pure bipartite quantum states $|\psi^{AB}\rangle$ we get two possibilities: either $|\psi^{AB}\rangle$ is a product state,
$|\psi^{AB}\rangle=|\phi^A\rangle\otimes|\varphi^B\rangle$,
for some $|\phi^A\rangle\in\mathscr{H}^A$ and $|\varphi^B\rangle\in\mathscr{H}^B$, which is separable and does not manifest any quantum correlations; or  $|\psi^{AB}\rangle$ is not factorizable, in which case it is entangled and hence manifests quantum correlations. This is the whole story as far as pure states are concerned: entanglement entirely captures every aspect of quantum correlations.

\subsection{\!\!Identifying  classically correlated states}

For bipartite quantum mixed states, however, the story becomes more complicated than that, as there are many paradigms that we can adopt in order to define what a classically correlated state is. One paradigm identifies the classically correlated states with the states that can be described by a local realistic model. According to this paradigm, only a restricted aristocracy of quantum states are not classically correlated, the so-called non-local states \cite{brunner2014bell}. Another paradigm is the one corresponding to entanglement, wherein classically correlated states can be written as convex combinations of tensor product of pure states, so-called separable states \cite{horodecki2009quantum}, i.e.,
\begin{equation}\label{Eq:separable}
\sigma^{AB}_{\text{sep}}=\sum_i p_i |\phi_i^{A}\rangle\langle\phi_i^{A}|\otimes|\varphi_i^{B}\rangle\langle\varphi_i^{B}|,
\end{equation}
with $\{p_i\}$ being a probability distribution, $|\phi_i^A\rangle\in\mathscr{H}^A$ and $|\varphi_i^B\rangle\in\mathscr{H}^B$. Separable states remind us of what happens in the classical setting, wherein all joint probability distributions can be written as a convex combination of products of Kronecker deltas, which are indeed the classical pure states. According to the entanglement paradigm, the right of being quantumly correlated is extended from the restricted aristocracy of non-local states to the broader bourgeoisie of non-separable quantum states. Finally, we get to the paradigm representing the focus of this manuscript, which goes even beyond entanglement, thus allowing the right of being quantumly correlated to almost all the population of quantum states.

As we have already mentioned, the embedding of a state of a classical system into the quantum state space is the corresponding classical mixture of elements of an orthonormal basis. However, when embedding the state of a classical composite system, imposing just the orthonormality of the basis is not enough, as one also needs to impose that such a basis is factorizable in order for the corresponding classical mixture to be entirely simulated by a classical bipartite state. This gives rise to  a so-called classical-classical state, i.e.,
\begin{equation}\label{Eq:classical-classical}
\chi_{\text{cc}}^{AB}=\sum_{i=1}^{d_A} \sum_{j=1}^{d_B} p_{ij}^{AB} |i^A\rangle\langle i^A|\otimes |j^B\rangle\langle j^B|,
\end{equation}
where ${\{p_{ij}^{AB}\}}_{i,j=1}^{d_A,d_B}$ is a joint probability distribution, while ${\{|i^A\rangle\}}_{i=1}^{d_A}$ and ${\{|j^B\rangle\}}_{j=1}^{d_B}$ are orthonormal bases of $\mathscr{H}^A$ and $\mathscr{H}^B$, respectively. One can indeed easily see that the marginal states of a classical-classical state are still classical states, i.e., classical mixtures of elements of an orthonormal basis: $\chi_{\text{cc}}^A = \mbox{Tr}_B(\chi_{\text{cc}}^{AB})=\sum_{i=1}^{d_A} p_i^A |i^A\rangle\langle i^A|$ and $\chi_{\text{cc}}^B = \mbox{Tr}_A(\chi_{\text{cc}}^{AB})=\sum_{j=1}^{d_B} p_i^B |j^B\rangle\langle j^B|$, where we have that ${\{p_i^A=\sum_{j=1}^{d_B} p_{ij}^{AB}\}}_{i=1}^{d_A}$ and ${\{p_j^B=\sum_{i=1}^{d_A} p_{ij}^{AB}\}}_{j=1}^{d_B}$ are exactly the marginal probability distributions of the joint probability distribution $\{p_{ij}^{AB}\}_{i,j=1}^{d_A,d_B}$.

Furthermore, one can also define the embedding of a classical state of only subsystem $A$ into the quantum state space of a bipartite quantum system $AB$ by considering what is known as a classical-quantum state, i.e.,
\begin{equation}\label{Eq:classical-quantum}
\chi_{\text{cq}}^{AB}=\sum_{i=1}^{d_A} p_{i}^{A} |i^A\rangle\langle i^A|\otimes \rho_i^B,
\end{equation}
with ${\{p_{i}^{A}\}}_{i=1}^{d_A}$ being a probability distribution, ${\{|i^A\rangle\}}_{i=1}^{d_A}$ an orthonormal basis of $\mathscr{H}^A$ and $\rho_i^B$ arbitrary states of subsystem $B$. In this case, one can easily see that in general only the marginal state of subsystem $A$ is still a classical state, while the marginal state of subsystem $B$ could be in principle any quantum state, i.e., $\chi_{\text{cq}}^A=\mbox{Tr}_B(\chi_{\text{cq}}^{AB}) = \sum_{i=1}^{d_A} p_i^A |i^A\rangle\langle i^A|$ while $\chi_{\text{cq}}^B=\mbox{Tr}_A(\chi_{\text{cq}}^{AB}) = \sum_{i=1}^{d_A} p_i^A \rho_i^B$.

An analogous definition holds when considering the embedding of a classical state of only subsystem $B$ into the state space of a bipartite quantum system $AB$, also called a quantum-classical state, i.e.,
\begin{equation}
\chi_{\text{qc}}^{AB}=\sum_{j=1}^{d_B} p_{j}^{B} \rho_j^A\otimes|j^B\rangle\langle j^B|,
\end{equation}
with ${\{p_{j}^{B}\}}_{j=1}^{d_B}$ being a probability distribution, ${\{|j^B\rangle\}}_{j=1}^{d_B}$ an orthonormal basis of $\mathscr{H}^B$ and $\rho_j^A$ arbitrary states of subsystem $A$.

Classical-classical, classical-quantum, and quantum-classical states, which we may collectively refer to as classically correlated states, form non-convex sets of measure zero and nowhere dense in the space of all bipartite quantum states $\rho^{AB}$~\cite{ferraro2010almost}. This is in stark contrast with the set of separable states, which is convex and occupies a finite volume in the state space instead \cite{horodecki2009quantum}.


\section{\!\!\!Quantifying quantum correlations: Quantum discord}

As mentioned in the introduction, and as will be shown in more detail in the following sections, quantum correlations beyond entanglement can represent a resource for some operational tasks and allow us to achieve them with an efficiency that is unreachable by any classical means. The quantification of this type of quantumness is thus necessary to gauge the quantum enhancement when performing such tasks.

Let us start from the quantification of quantum correlations for pure states. We have already mentioned that in this case the entire amount of quantum correlations is captured by entanglement. This can be in turn described by the fact that, when dealing with pure bipartite quantum states that are not factorizable, the best possible knowledge of a whole does not include the best possible knowledge of all its parts, as the corresponding marginal states are necessarily mixed. Such a loss of information on the pure state of the whole system when accessing only part of it, as quantified e.g. by the von Neumann entropy of any of the marginal states, captures exactly the entanglement, and thus the whole quantum correlations, between the two parties\footnote{Note that the reduced states $\rho^A$ and $\rho^B$ of any bipartite pure state have the same eigenvalues and so the same von Neumann entropy, thus making the definition of the entropy of entanglement $E_\mathcal{S}$ well posed.}:
\begin{equation}\label{Eq:EntropyE}
E_\mathcal{S}(|\psi^{AB}\rangle)=\mathcal{S}(\rho^A)=\mathcal{S}(\rho^B).
\end{equation}
The pure state entanglement quantifier $E_\mathcal{S}$ is also known as \textit{entropy of entanglement}.

Let us now move on to the quantification of quantum correlations beyond entanglement for mixed states. Both adopting an entropic viewpoint and a thorough comparison with the classical setting will turn out to be crucial at this stage, as happened in the previous section when addressing the characterization of quantum correlations. When a bipartite classical system $AB$ is in a mixed state $\textbf{p}^{AB}$, then we have some ignorance about it that can be quantified by its strictly positive Shannon entropy $\mathcal{S}(\textbf{p}^{AB})$. At the same time, quite intuitively, it turns out that the overall ignorance about the marginal states $\textbf{p}^A$ and $\textbf{p}^B$ of the two subsystems $A$ and $B$ treated separately, which is quantified by the quantity $\mathcal{S}(\textbf{p}^A) + \mathcal{S}(\textbf{p}^B)$, is necessarily higher than or equal to the ignorance about the state of the combined bipartite system, which is instead quantified by $\mathcal{S}(\textbf{p}^{AB})$. In other words, there is in general a loss of information on the state of the whole system when accessing only its parts. This can be quantified by the so-called \textit{mutual information}:
\begin{equation}\label{Eq:mutualI}
\mathcal{I}(\textbf{p}^{AB}) =\mathcal{S}(\textbf{p}^A) + \mathcal{S}(\textbf{p}^B) - \mathcal{S}(\textbf{p}^{AB}).
\end{equation}
Such a loss of information when accessing a composite system locally is attributed to underlying correlations between the subsystems, so that the mutual information stands as a fully fledged quantifier of correlations. We can think of two correlated subsystems $A$ and $B$ as two accomplices. If the policemen interrogate them separately, the more the two accomplices are correlated, the less information the policemen will manage to gain regarding what $AB$ did together, with their mutual information representing exactly the amount of information that the two accomplices are hiding to the policemen. Clearly, for pure bipartite classical states we always get a zero mutual information, as both the composite system state and the marginal states are pure and so their Shannon entropies are all zero and there is no loss of information in accessing the composite system locally. This entails that it is impossible to have correlations between classical systems sharing a pure state, contrary to what happens within the quantum world where we can have entanglement for pure states. More generally, the mutual information is equal to zero if, and only if, the bipartite classical state $\textbf{p}^{AB}$ is factorizable, i.e., $p_{ij}^{AB}=p_i^Ap_j^B$ for any $i$ and $j$, which is indeed the paradigmatic form of probability distribution that does not manifest any correlations at all.

Yet there is another equivalent perspective from which we can look at correlations in the classical setting. Let us first define $\textbf{p}^{A|B=j}={\{p^{A|B=j}_i\}}_{i=1}^{d_A} = {\{p^{AB}_{ij}/p^B_j\}}_{i=1}^{d_A}$ as the conditional probability distribution of subsystem $A$ after we know that subsystem $B$ occupies exactly the point $j$ of its phase space. Analogously, we define $\textbf{p}^{B|A=i}={\{p^{B|A=i}_j\}}_{j=1}^{d_B} = {\{p^{AB}_{ij}/p^A_i\}}_{j=1}^{d_B}$ as the conditional probability distribution of subsystem $B$ after we know that subsystem $A$ occupies exactly the point $i$ of its phase space. Then, one can prove that the mutual information of the bipartite state $\textbf{p}^{AB}$ is equal to the following quantity:
\begin{equation}\label{Eq:mutualJ}
\mathcal{J}(\textbf{p}^{AB}) = \mathcal{S}(\textbf{p}^A) - \sum_{j=1}^{d_B} p^B_j \mathcal{S}(\textbf{p}^{A|B=j})  = \mathcal{S}(\textbf{p}^B) - \sum_{i=1}^{d_A} p^A_i \mathcal{S}(\textbf{p}^{B|A=i}).
\end{equation}

The above equivalent expressions of the mutual information tell us that the more two subsystems $A$ and $B$ are correlated, the more the ignorance about one subsystem decreases on average when we know the state of the other subsystem. On the other hand, if $A$ and $B$ are not correlated at all, then gaining some information about one subsystem does not help us in gaining any information about the other subsystem.

Now the question is: how can we translate such a machinery into the quantum setting in order to quantify quantum correlations beyond entanglement? Clearly, we can start by defining the \textit{quantum mutual information} in order to quantify the totality of correlations of bipartite quantum states $\rho^{AB}$ as follows:
\begin{equation}
\mathcal{I}(\rho^{AB}) = \mathcal{S}(\rho^A) + \mathcal{S}(\rho^B) - \mathcal{S}(\rho^{AB}),
\end{equation}
where $\mathcal{S}$ here denotes the von Neumann entropy.
In analogy with the classical case, the quantum mutual information is equal to zero if, and only if, $\rho^{AB}$ is factorizable, i.e., $\rho^{AB}=\rho^A\otimes\rho^B$ and thus there are no correlations whatsoever, not even classical ones, between $A$ and $B$. However, in order to fully answer our question we need to find out how to discern the portion of the total correlations that is purely quantum from the one that can be regarded as mere classical correlations, a problem that was rigorously addressed for the first time by Henderson and Vedral in \cite{henderson2001classical}.

To this purpose it will be crucial to translate in the quantum setting also the quantity $\mathcal{J}$, which in the classical setting represents just an equivalent expression for the mutual information. We thus need to define also in the quantum setting the conditional state of one subsystem given that we have gained some information about the other subsystem. The most intuitive way to gain information about a single quantum subsystem, say $A$, is to measure a local observable of the form $O^A\otimes \mathbb{I}^B$, where $O^A$ is a Hermitian operator on $\mathscr{H}^A$ while $\mathbb{I}^B$ is the identity operator on $\mathscr{H}^B$. As we have already mentioned, the result of such a measurement is in general uncertain and can map the system, with probability $p_i^A=\mbox{Tr}[(\Pi_i^A\otimes\mathbb{I}^B)\rho^{AB}]$, into the state $\rho^{AB|A=i}_{\boldsymbol{\Pi}^A}= (\Pi_i^A\otimes\mathbb{I}^B)\rho^{AB}(\Pi_i^A\otimes\mathbb{I}^B)/p_i^A$, where the rank-one projectors $\boldsymbol{\Pi}^A=\{\Pi_i^A\}$ are the eigenstates of $O^A$. Therefore, the conditional state of subsystem $B$ after such a local measurement has been performed on $A$ and the result $i$ has been obtained is $\rho^{B|A=i}_{\boldsymbol{\Pi}^A} = \mbox{Tr}_A(\rho^{AB|A=i}_{\boldsymbol{\Pi}^A})$.  We can thus define the decrease on average of the entropy of $B$ given that we have performed the local measurement on $A$ described by the rank-one projectors $\boldsymbol{\Pi}^A$ on $A$ as
\begin{equation}
\mathcal{J}_{\boldsymbol{\Pi}^A}(\rho^{AB})= \mathcal{S}(\rho^B) - \sum_{i} p^A_i \mathcal{S}(\rho^{B|A=i}_{\boldsymbol{\Pi}^A}).
\end{equation}
Some remarks are now in order. Firstly, contrary to the classical case, we can define different versions of conditional states $\rho^{B|A=i}_{\boldsymbol{\Pi}^A}$ of $B$ given $A$, and so different versions of the quantity $\mathcal{J}_{\boldsymbol{\Pi}^A}$, just by varying the local measurement $\boldsymbol{\Pi}^A$ that has been performed on $A$. Secondly, one can even consider more general kinds of local measurements (described by positive operator-valued measures), but we restrict to rank-one projective measurements here  for the sake of simplicity.

Finally, the correlations underlying such a gain of information about subsystem $B$, when accessing locally  subsystem $A$ after the local measurement $\boldsymbol{\Pi}^A$,  can be considered classical from the perspective of subsystem $A$, as they are nothing but the correlations that are left into the post-measurement state $\Pi^A[\rho^{AB}]=\sum_i p_i^A \rho^{AB|A=i}_{\boldsymbol{\Pi}^A}$, which is clearly a classical-quantum state. In other words, one can see that the following equality holds:
\begin{equation}
\mathcal{J}_{\boldsymbol{\Pi}^A}(\rho^{AB})=\mathcal{I}\left(\Pi^A[\rho^{AB}]\right). \\
\end{equation}

Therefore, if one wants to extract from the total correlations $\mathcal{I}(\rho^{AB})$ of the bipartite state $\rho^{AB}$ the purely quantum portion of correlations from the perspective of subsystem $A$, i.e., the amount of mutual information of $A$ and $B$ that can be never classically extracted via a local measurement on $A$, not even by performing a maximally informative one, then one can consider the following quantity:
\begin{equation}\label{Eq:DiscordA}
Q_\mathcal{I}^A(\rho^{AB}) = \mathcal{I}(\rho^{AB}) - \max_{\boldsymbol{\Pi}^A} \mathcal{J}_{\boldsymbol{\Pi}^A}(\rho^{AB}),
\end{equation}
where the maximization is over all rank-one local projective measurements on $A$. $Q_\mathcal{I}^A $ is the celebrated quantifier of quantum correlations beyond entanglement from the perspective of subsystem $A$ that goes under the name of \textit{quantum discord} and was introduced by Ollivier and Zurek in \cite{ollivier2001quantum}.
The complementary quantity
\begin{equation}
\mathcal{J}^A(\rho^{AB}) = \max_{\boldsymbol{\Pi}^A} \mathcal{J}_{\boldsymbol{\Pi}^A}(\rho^{AB}),
\end{equation}
quantifies the classical correlations from the perspective of subsystem $A$ as formalized by Henderson and Vedral in \cite{henderson2001classical}. In this way, quantum discord $Q_\mathcal{I}^A(\rho^{AB})$ and classical correlations $\mathcal{J}^A(\rho^{AB})$ add up to the total correlations quantified by the mutual information  $\mathcal{I}(\rho^{AB})$, and we have addressed the original question posed in this section, by finding a meaningful way to separate the quantum from the classical portion of correlations in a state $\rho^{AB}$, from the perspective of subsystem $A$.

Analogous definitions hold when measuring locally subsystem $B$, by swapping the roles of $A$ and $B$. In particular, the quantum discord from the perspective of subsystem $B$ can be  defined as:
\begin{equation}\label{Eq:DiscordB}
Q_\mathcal{I}^B(\rho^{AB}) = \mathcal{I}(\rho^{AB}) - \max_{\boldsymbol{\Pi}^B} \mathcal{J}_{\boldsymbol{\Pi}^B}(\rho^{AB}),
\end{equation}
where the maximization is over all rank-one local projective measurements on $B$.

A further couple of remarks are  in order before concluding this section. Firstly, a fundamental asymmetry arises between how the quantum correlations between $A$ and $B$ are perceived by each subsystem, due to the fact that in general $Q_\mathcal{I}^A(\rho^{AB})$ is different from $Q_\mathcal{I}^B(\rho^{AB})$. Quantum discord is, in fact, a one-sided measure of quantumness of correlations. However, such an asymmetry can be bypassed by considering the action of local joint measurements on both $A$ and $B$, and defining accordingly symmetric (or two-sided) quantifiers of quantum and classical correlations from the perspective of either $A$ or $B$ within the same entropic framework adopted in this section \cite{piani2008no,wu2009correlations,divincenzo2004locking}. More details on these quantifiers, that may be denoted respectively by $Q_{\mathcal{I}}^{AB}(\rho_{AB})$ and  $\mathcal{J}^{AB}(\rho^{AB})$, as well as their interplay with one-sided measures, are available in \cite{lang2011entropic,adesso2016measures}.

Secondly, by using both the fact that classical-quantum states $\chi^{AB}_{\text{cq}}$ can be left invariant by at least one local projective measurement $\boldsymbol{\Pi}^A$ on $A$, i.e., $\Pi^A[\chi^{AB}_{\text{cq}}]=\chi^{AB}_{\text{cq}}$, and the fact that the result of such a measurement applied to any state is always a classical-quantum state,\footnote{Proving these statements can be left as an exercise to the reader.} i.e., $\Pi^A[\rho^{AB}]=\chi^{AB}_{\text{cq}}$ for any $\rho^{AB}$, one can easily show that $Q_\mathcal{I}^A(\rho^{AB})=0$ if, and only if, $\rho^{AB}$ is classical-quantum. An analogous result holds for quantum correlations with respect to $B$, i.e., $Q_\mathcal{I}^B(\rho^{AB})=0$ if, and only if, $\rho^{AB}$ is quantum-classical.
This cements the paradigm adopted in this manuscript, according to which almost all quantum bipartite states, and not only entangled states, manifest genuinely quantum features that can be attributed to non-classical correlations.

\section{\!\!\!Interpreting quantum correlations: Local broadcasting} \label{Sec:LBRB}

We have identified the classically and quantumly correlated states and provided an entropic way to measure quantum correlations in terms of the discord. It is now time for us to place what we have learnt in more concrete terms by understanding the role of quantum correlations in an operational task:  local broadcasting~\cite{piani2008no,luo2010quantum}.

Let us first consider copying of information. This happens all the time in the classical realm: from hard drives to mobile telephones -- our modern world relies on the ability to freely copy information. In stark contrast, general copying of information is expressly prohibited in quantum mechanics by the no-broadcasting theorem~\cite{barnum1996noncommuting}, which is a generalization of the well-known no-cloning theorem~\cite{wootters1982single,dieks1982communication}. Think of a quantum system $A$  in one of two states $\rho_{1}$ or $\rho_{2}$. We attach an ancilla $A'$ in the state $\sigma$ to get the composite state $\rho_{i}^A \otimes \sigma^{A'}$ with $i \in \{1,2\}$. The goal is to perform some transformation $\mathcal{E}$ to the composite state to get $\tilde{\rho}_i^{AA'}=\mathcal{E}[\rho_{i}^A \otimes \sigma^{A'}]$ such that  $\mbox{Tr}_{A'}(\tilde{\rho}_i^{AA'}) = \mbox{Tr}_{A}(\tilde{\rho}_i^{AA'}) = \rho_{i}$ for both $i=1,2$. 
In other words, we want to be able to copy two arbitrary quantum states $\rho_{1}$ and $\rho_{2}$. However, it turns out this is only possible if $\rho_{1}$ and $\rho_{2}$ commute, which effectively reduces to copying of classical information.

The objective of local broadcasting is similar~\cite{piani2008no}. Consider now a composite state $\rho^{AB}$ shared between two subsystems $A$ and $B$. We give each subsystem an ancilla $A'$ and $B'$ so that the joint state is $\rho^{AB}\otimes \ket{0^{A'}}\bra{0^{A'}}\otimes \ket{0^{B'}}\bra{0^{B'}}$ and ask if there exists a local operation $\mathcal{E}^{AA'} \otimes \mathcal{E}^{BB'}$ so that we get the state
$\tilde{\rho}^{AA'BB'} = (\mathcal{E}^{AA'} \otimes \mathcal{E}^{BB'})\big[\rho^{AB}\otimes \ket{0^{A'}}\bra{0^{A'}}\otimes \ket{0^{B'}}\bra{0^{B'}}\big]$
obeying the relation
$\mbox{Tr}_{A'B'}(\tilde\rho^{AA'BB'}) = \mbox{Tr}_{AB}(\rho^{AA'BB'}) = \rho^{AB}$.
More generally, we can consider the task of simply distributing the (total) \emph{correlations} $\mathcal{I}(\rho^{AB})$ of $\rho^{AB}$, and ask if there are local operations such that
$\mathcal{I}\big(\mbox{Tr}_{A'B'}(\tilde\rho^{AA'BB'})\big) = \mathcal{I}\big(\mbox{Tr}_{AB}(\tilde\rho^{AA'BB'})\big) = \mathcal{I}\big(\rho^{AB}\big)$.
This is what we mean by local broadcasting, and it was shown in~\cite{piani2008no} that such a process can only take place perfectly if $\rho^{AB}$ is classical-classical, otherwise we lose correlations during our attempt at local broadcasting.

A similar one-sided version of local broadcasting has also been proposed in~\cite{luo2010quantum}. Here, we just give subsystem $A$ their ancilla $A'$ and ask if there is a local operation $\mathcal{E}^{AA'}\otimes \mathbb{I}^{B}$ yielding $\tilde\rho^{AA'B} = (\mathcal{E}^{AA'}\otimes \mathbb{I}^{B}) [\rho^{AB}\otimes \ket{0^{A'}}\bra{0^{A'}}]$ such that $\mathcal{I}(\mbox{Tr}_{A'}(\tilde\rho^{AA'B})) = \mathcal{I}(\mbox{Tr}_{A}(\tilde\rho^{AA'B})) = \mathcal{I}(\rho^{AB})$. As you might have guessed, this version of local broadcasting can occur only if $\rho^{AB}$ is classical-quantum.

\begin{figure*}
\includegraphics[width=\textwidth]{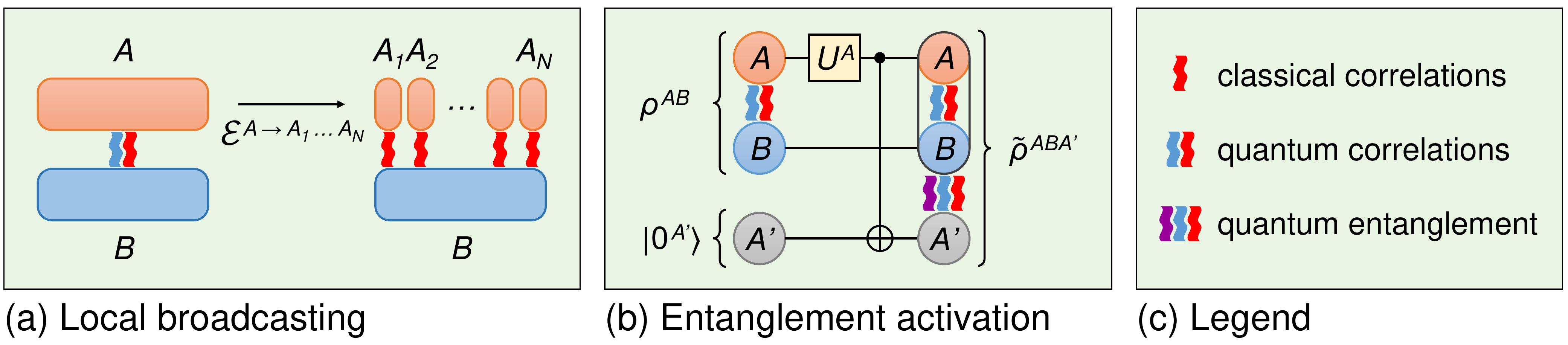} \vspace*{-.3cm}
\caption{\label{Fig:Discillusioned}
{\textsf{\bfseries Operational interpretations and quantification of quantum correlations.}} \\[1pt] 
{\textsf{\small (a) Local broadcasting of correlations [Section~\ref{Sec:LBRB}].
Two quantum systems $A$ and $B$ are initially in an arbitrary bipartite state $\rho^{AB}$ with generally classical and quantum correlations.
If a local channel $\mathcal{E}^{A \rightarrow A_{1} \ldots A_{N}}$ is applied to $A$ which redistributes it
into asymptotically many fragments $A_1, \ldots, A_N$, then the only correlations remaining on average between each
fragment $A_i$ and subsystem $B$ are classical ones, while quantum correlations, quantified by the quantum discord $Q_{\cal I}^A(\rho_{AB})$,
cannot be shared. This can be seen as a manifestation of {\em quantum Darwinism} \cite{brandao2015generic}.
(b) Scheme of a pre-measurement interaction acting on subsystem $A$ of a bipartite system $AB$, described as a local unitary $U^A$ on $A$, followed by a generalized control-\textsc{NOT} operation with an ancilla $A'$ (which plays the role of a measurement apparatus). Provided $A'$ is initialized in a pure state $\ket{0^{A'}}$, the output pre-measurement state $\tilde{\rho}^{ABA'}$ is always entangled along the $AB:A'$ split if and only if the initial
state $\rho^{AB}$ of the system is not classical-quantum, i.e., contains general quantum correlations from the perspective of subsystem $A$ \cite{streltsov2011linking,piani2011all}. The minimum entanglement $E^{AB:A'}(\tilde{\rho}^{ABA'})$ between $AB$ and $A'$ in the pre-measurement state, where the minimization is over all the local bases on $A$ specified by $U^A$, quantifies the quantum correlations $Q^A_E(\rho^{AB})$ in the input bipartite state $\rho^{AB}$, according to the entanglement activation paradigm [Section~\ref{Sec:EASports}].  (c) Graphical legend for the different types of correlations appearing in panels (a) and (b)}.}}
\end{figure*}

We thus have a very intuitive characterization of classical-classical states and classical-quantum states: they are exactly the states that can be locally broadcast. So can we use this concept of local broadcasting to \emph{quantify} the quantum correlations present in a state? Now let us imagine that $A$ wants to distribute their correlations with $B$ to $N$ ancillae $\{A_{i}\}_{i=1}^{N}$ using local operations $\mathcal{E}^{A \rightarrow A_{1} \ldots A_{N}}$~\cite{brandao2015generic}. If we define the reduced state of each pair $A_{i}$ and $B$ after such local operations as
\begin{equation}
\tilde\rho^{A_i B} = \mbox{Tr}_{\otimes_{j \neq i} A_j} \big\{(\mathcal{E}^{A \rightarrow A_1\ldots A_N} \otimes \mathbb{I}^B)[\rho^{AB}] \big\} \, ,
\end{equation}
we know from the above analysis that correlations will never increase, i.e., $\mathcal{I}(\tilde\rho^{A_{i}B}) \leq \mathcal{I}(\rho^{AB})$, with equality only if $\rho^{AB}$ is classical-quantum. Let us suppose that $\rho^{AB}$ is not classical-quantum, but we want to distribute our correlations in an efficient way, i.e., losing the least possible amount of correlations. We can consider the loss of correlations ${\cal I}(\rho^{AB}) - {\cal I}(\tilde\rho^{A_iB})$ for each ancilla. Averaging this quantity over all ancillae then gives a good figure of merit for our redistribution of correlations. By further minimizing this figure of merit over all possible local operations, we get
\begin{equation}\label{Eq:Tuco}
\overline{\Delta}^A_{(N)}(\rho^{AB}) = \min_{\mathcal{E}^{A \rightarrow A_1\ldots A_N}} \frac{1}{N} \sum_{i=1}^N \big[{\cal I}(\rho^{AB}) - {\cal I}(\tilde\rho^{A_iB})\big].
\end{equation}
This quantity is zero if $\rho^{AB}$ is classical-quantum, and positive otherwise. Can its value quantify the quantum correlations of $\rho^{AB}$? Remarkably, in the limit of infinitely many ancillae, it has been proven in \cite{brandao2015generic} that the quantity in Eq.~(\ref{Eq:Tuco}) reproduces exactly the quantum discord given by Eq.~(\ref{Eq:DiscordA}) [see Figure~\ref{Fig:Discillusioned}(a)]:
\begin{equation}\label{Eq:NoDoze}
\lim_{N \rightarrow \infty} \overline{\Delta}^A_{(N)}(\rho^{AB}) = Q^{A}_{\mathcal{I}}(\rho^{AB}) \, .
\end{equation}
This relation provides a striking operational understanding of quantum discord as the minimum average loss of correlations if one attempts to redistribute the correlations between $A$ and $B$ in  the state  $\rho^{AB}$ to infinitely many ancillae on $A$'s side: paraphrasing the words of~\cite{streltsov2013quantum}, ``quantum correlations cannot be shared''. We note that additional operational interpretations for the quantum discord in quantum information theory and thermodynamics have been discovered, as reviewed in \cite{modi2012classical,streltsov2015quantum,adesso2016measures}.

\section{\!\!\!Alternative characterizations of quantum correlations}

So far we have focused on the characterization of classically correlated states and the quantification of quantum correlations in an entropic setting, using the quantum discord. One property that we have pointed out along the way is that the classically correlated states are insensitive to a local complete rank-one projective measurement, a hallmark feature of the classical world. It has also been shown that classically correlated states are the only ones that are locally broadcastable, another intuitive property arising from the inability to copy general quantum states. It turns out that there is a whole raft of equivalent defining properties for the classically correlated states, and that with each property comes another way to quantify the quantum correlations \cite{adesso2016measures}. The quantum discord accounts for the loss of correlations due to local measurements, but it is just one of many ways to measure the quantum correlations of a state. We will outline two more key properties of classically correlated states in the following, along with the corresponding method of measuring quantum correlations.

\renewcommand{\baselinestretch}{0.95}

\renewcommand{\baselinestretch}{1.05}

\subsection{\!\!Local coherence}

Recall that we define the incoherent states with respect to a reference basis $\{\ket{i}\}_{i=1}^{d}$ as those diagonal in this basis, i.e., states that can be written as
$\delta = \sum_{i=1}^{d} p_{i} \ket{i}\bra{i}$
for some probability distribution $\{p_{i}\}_{i=1}^{d}$. Any state that is not diagonal in this basis is called coherent \cite{baumgratz2014quantifying,streltsov2016review}. Now let us consider a bipartite quantum system $AB$ with local reference bases $\{\ket{i^{A}}\}_{i=1}^{d_{A}}$ in $A$ and $\{\ket{j^{B}}\}_{j=1}^{d_{B}}$ in $B$. States incoherent with respect to the product basis $\{\ket{i^{A}} \otimes \ket{j^{B}}\}_{i,j=1}^{d_{A},d_{B}}$ can be written as
\begin{equation}
\delta_{\text{ii}}^{AB} = \sum_{i=1}^{d_{A}} \sum_{j=1}^{d_{B}} p_{ij}^{AB} \ket{i^{A}}\bra{i^{A}} \otimes \ket{j^{B}}\bra{j^{B}}
\end{equation}
for some joint probability distribution $\{p_{ij}^{AB}\}$, while states incoherent in the local reference basis $\{\ket{i^{A}}\}_{i=1}^{d_{A}}$ are written as
\begin{equation}\label{Eq:Incoherent-Quantum}
\delta^{AB}_{\text{iq}} = \sum_{i=1}^{d_{A}} p_{i}^{A} \ket{i^{A}}\bra{i^{A}} \otimes \rho_{i}^{B}
\end{equation}
for some probability distribution $\{p_{i}^{A}\}$ and with arbitrary states $\rho_{i}^{B}$ of subsystem $B$. We can say that these locally incoherent states are incoherent-incoherent and incoherent-quantum, respectively. Take a look back at Eqs.~(\ref{Eq:classical-classical}) and (\ref{Eq:classical-quantum}) describing the classically correlated states. You would be forgiven for thinking that they are identical to the equations given above! However, there is a subtlety here: the locally incoherent states are diagonal in a \emph{fixed} local basis, while the classically correlated states are diagonal in \emph{some} local basis. This analogy then provides us with another characterization of the classically correlated states, i.e. classical-classical states are incoherent-incoherent for some product basis on $A$ and $B$, while classical-quantum states are incoherent-quantum for some local basis on $A$~\cite{adesso2016measures}.

On the other hand, quantumly correlated states are coherent in every local basis. Can we then use measures of coherence to inform us on the amount of quantum correlations? Consider the observable $K = \sum_{i=1}^{d} k_{i} \ket{i}\bra{i}$ diagonal in a fixed reference basis $\{\ket{i}\}_{i=1}^{d}$. One way to measure the coherence of a state $\rho$ with respect to the reference basis, or more precisely its asymmetry with respect to translations generated by the observable $K$, is by means of the quantum Fisher information~$F(\rho,K)$~\cite{girolami2014observable,marvian2014extending}. This quantity plays a fundamental role in quantum metrology~\cite{braunstein1994statistical} and indicates the ultimate precision achievable using a quantum probe state $\rho$ to estimate a parameter encoded in a unitary evolution generated by the observable $K$. Let us now fix a family of \emph{local} observables $K_{\Gamma}^{A}= \sum_{i=1}^{d} k_{i}^{A} \ket{i^{A}}\bra{i^{A}}$ on subsystem $A$ with fixed non-degenerate spectrum $\Gamma = \{k_{i}^{A}\}_{i=1}^{d_{A}}$. Defining the minimum of $F(\rho^{AB} ,K_{\Gamma}^{A} \otimes \mathbb{I}^{B})$ over all local observables $K_{\Gamma}^{A}$ with spectrum $\Gamma$ gives a measure of quantum correlations~\cite{girolami2014quantum}:
\begin{equation}\label{Eq:IP}
Q^{A}_{F} (\rho^{AB}) = \frac14 \inf_{K_{\Gamma}^{A}} F(\rho^{AB} ,K_{\Gamma}^{A} \otimes \mathbb{I}^{B}).
\end{equation}
Such a measure embodies the worst-case scenario sensitivity of a bipartite state $\rho^{AB}$ when a parameter is imprinted onto subsystem $A$ by any of the observables $K_{\Gamma}^{A}$: a  process that is fundamentally linked to quantum interferometry and hence motivates the naming of $Q^{A}_{F} (\rho^{AB})$ as the {\it interferometric power}~\cite{girolami2014quantum}. While there are many other good measures of quantum coherence~\cite{streltsov2016review}, from which one can define corresponding measures of quantum correlations (by minimization over local reference bases)~\cite{adesso2016measures}, the interferometric power is one of the most compelling as it brings together quantum coherence, quantum correlations, and  metrology. Another advantage of this measure is that $Q^{A}_{F} (\rho^{AB})$ admits a computable formula for any state $\rho^{AB}$ whenever $A$ is a qubit \cite{girolami2014quantum}, while no such analytical formula is presently available for the quantum discord $Q^{A}_{\cal I} (\rho^{AB})$ of general two-qubit or qubit-qudit states.

\subsection{\!\!Entanglement activation}\label{Sec:EASports}

Let us now examine more closely the workings of a local projective measurement $\Pi^{A}[\rho^{AB}] = \sum_{i=1}^{d_{A}} (\Pi^A_i \otimes \mathbb{I}^{B}) \rho^{AB} (\Pi^A_i \otimes \mathbb{I}^{B})$ with local projectors $\Pi^A_i=\ket{i^{A}}\bra{i^{A}}$ acting on subsystem $A$ of a bipartite state $\rho^{AB}$. According to von Neumann's model~\cite{vonneumann1932mathematical}, this measurement can be realized in two steps. First, subsystem $A$ is allowed to interact with an ancilla $A'$, initialized in a fiducial pure state $\ket{0^{A'}}\bra{0^{A'}}$, through a unitary $U^{AA'}_{\{\ket{i^{A}}\}}$. The unitary acts in the following way
\begin{equation}
U^{AA'}_{\{\ket{i^{A}}\}} \ket{i^{A}} \otimes \ket{0^{A'}} = \ket{i^{A}} \otimes \ket{i^{A'}},
\end{equation}
and can be realized by the combination $U^{AA'}_{\{\ket{i^{A}}\}} = C^{AA'}(U^{A}_{\{\ket{i^{A}}\}} \otimes \mathbb{I}^{A'})$ of a local unitary $U^{A}_{\{\ket{i^{A}}\}}$ which sets the basis of measurement, followed by a generalized controlled-\textsc{not} gate $C^{AA'}$, whose action on the computational basis $\ket{i^{A}} \otimes \ket{j^{A'}}$ of $\mathbb{C}^d \otimes \mathbb{C}^d$ is  $C^{AA'} \ket{i^{A}} \otimes \ket{j^{A'}} = \ket{i^{A}} \otimes \ket{i \oplus j^{A'}}$, with  $\oplus$ denoting addition modulo $d$. 
 The resultant state after applying the unitary $U^{AA'}_{\{\ket{i^{A}}\}}$ to $A$, $A'$ and $B$ is
\begin{equation}
\tilde{\rho}^{ABA'}_{\{\ket{i^{A}}\}} = (U^{AA'}_{\{\ket{i^{A}}\}} \otimes \mathbb{I}^{B})(\rho^{AB} \otimes \ket{0^{A'}}\bra{0^{A'}})(U^{AA'}_{\{\ket{i^{A}}\}} \otimes \mathbb{I}^{B})^{\dagger},
\end{equation}
which is known as the pre-measurement state. Next, the local projective measurement is completed by partial tracing over subsystem $A'$, which is achieved by a readout of the ancilla $A'$ in its eigenbasis, so that $\mbox{Tr}_{A'} \tilde{\rho}^{ABA'}_{\{\ket{i^{A}}\}} = \Pi^{A}[\rho^{AB}]$.

During this process the ancilla $A'$ can become entangled with $A$ and $B$ due to the unitary $U^{AA'}_{\{\ket{i^{A}}\}}$. which means that the pre-measurement state $\tilde{\rho}^{ABA'}_{\{\ket{i^{A}}\}}$ may not be  separable along the bipartition $AB:A'$. However, sometimes  $\tilde{\rho}^{ABA'}_{\{\ket{i^{A}}\}}$ remains  separable along such cut. It turns out this is the case only when $\rho^{AB}$ is initially incoherent-quantum, of the form in Eq.~(\ref{Eq:Incoherent-Quantum}). It thus becomes clear that we can characterize the classical-quantum states of Eq.~(\ref{Eq:classical-quantum}) as exactly all and only the states for which there exists a local basis $\{\ket{i^{A}}\}$ such that the pre-measurement state $\tilde{\rho}^{ABA'}_{\{\ket{i^{A}}\}}$ is separable along the split $AB:A'$ \cite{streltsov2011linking}.

Similarly, if we consider a local projective measurement $(\Pi^{A} \otimes \Pi^{B})[\rho^{AB}]$ on both $A$ and $B$ in the bases $\{\ket{i^{A}}\}_{i=1}^{d_{A}}$ and $\{\ket{j^{B}}\}_{j=1}^{d_{B}}$, we can also introduce an ancilla $B'$ for $B$ and a corresponding pre-measurement state $\tilde{\rho}^{ABA'B'}_{\{\ket{i^{A}},\ket{j^{B}}\}}$. A similar line of thought can then be applied whereby we find that the classical-classical states of Eq.~(\ref{Eq:classical-classical}) are all and only the states for which the pre-measurement state is separable along the split $AB:A'B'$ for some local bases $\{\ket{i^{A}}\}_{i=1}^{d_{A}}$ and $\{\ket{j^{B}}\}_{j=1}^{d_{B}}$~\cite{piani2011all}.

From this analysis, it can be said that the classical correlations are not always {\em activated} into entanglement during a pre-measurement, while the quantum correlations always are. Such a conversion of non-classical resources due to a pre-measurement interaction has been demonstrated experimentally in \cite{adesso2014experimental}. Naturally, one can then aim to quantify the quantum correlations of $\rho^{AB}$ by measuring the entanglement of the corresponding pre-measurement state, via some chosen entanglement measure $E$, minimized over all local bases.

For every suitable $E$, we can then define a corresponding  (one-sided or two-sided) measure of quantum correlations~\cite{streltsov2011linking,piani2011all} as follows [see Figure~\ref{Fig:Discillusioned}(b)]
\begin{equation}\label{Eq:QCActivation}
\begin{aligned}
Q^{A}_{E} (\rho^{AB}) &= \inf_{\{\ket{i^{A}}\}} E^{AB:A'}(\tilde{\rho}^{ABA'}_{\{\ket{i^{A}}\}}),  \\
Q^{AB}_{E} (\rho^{AB}) &= \inf_{\{\ket{i^{A}},\ket{j^{B}}\}} E^{AB:A'B'}(\tilde{\rho}^{ABA'B'}_{\{\ket{i^{A}},\ket{j^{B}}\}}).
\end{aligned}
\end{equation}
One of the most remarkable features of this approach is that the measures so defined capture quantitatively the hierarchy of quantum correlations, as one has $Q^{AB}_{E} (\rho^{AB}) \geq Q^{A}_{E} (\rho^{AB}) \geq E^{A:B}(\rho^{AB})$ for any valid entanglement measure $E$ and any bipartite state $\rho^{AB}$, with equalities on pure states $\rho^{AB} = \ket {\psi^{AB}}\bra{\psi^{AB}}$.

For instance, one may choose the relative entropy of entanglement \cite{vedral1997quantifying}
\begin{equation}
E_{\text{R}}^{A:B}(\rho^{AB}) = \inf_{\sigma^{AB}_{\text{sep}}}   \mathcal{S}(\rho^{AB}||\sigma^{AB}_{\text{sep}}),
\end{equation}
 as our entanglement measure, where $\mathcal{S}(\rho||\sigma) = \mbox{Tr}(\rho \log \rho - \rho \log \sigma)$ is the relative entropy and  the minimization is over all separable states $\sigma^{AB}_{\text{sep}}$ of the form in Eq.~(\ref{Eq:separable}). The corresponding measures of quantum correlations, obtained by specifying $E$ as $E_{\text{R}}$ in Eqs.~(\ref{Eq:QCActivation}), are known respectively as relative entropy of discord (one-sided) and relative entropy of quantumness (two-sided). Interestingly, these measures have been proven equivalent to the following expressions~\cite{streltsov2011linking,piani2011all}:
\begin{equation}\label{Eq:DistanceBased}
Q^{A}_{E_{\text{R}}} (\rho^{AB}) = \inf_{\chi^{AB}_{\text{cq}}} \mathcal{S}(\rho^{AB}||\chi^{AB}_{\text{cq}}), \qquad
Q^{AB}_{E_{\text{R}}} (\rho^{AB}) = \inf_{\chi^{AB}_{\text{cc}}} \mathcal{S}(\rho^{AB}||\chi^{AB}_{\text{cc}}),
\end{equation}
with minimizations over the classical-quantum states of Eq.~(\ref{Eq:classical-quantum}) and the classical-classical states of Eq.~(\ref{Eq:classical-classical}), respectively.
This enriches the quantification of quantum correlations as potential resources for entanglement creation, with an additional geometric interpretation in terms of the distance\footnote{Note that the relative entropy is not strictly a distance because it is not symmetric in its arguments.} from the set(s) of classically correlated states. In turn, such a geometric approach can be used {\it a priori} to quantify quantum correlations adopting different distance functionals, as reviewed in \cite{spehner2014quantum,roga2015geometric,adesso2016measures}.


\section{\!\!\!Desiderata for measures of quantum correlations}

We have identified several alternative yet equivalent characterizations of the classically correlated states, in particular providing links with other fundamental elements of quantum mechanics such as coherence \cite{streltsov2016review} and entanglement \cite{horodecki2009quantum}. With each characterization of the classically correlated states comes another way to measure quantum correlations. Given such a catalogue of measures \cite{adesso2016measures}, it is sensible to wonder \emph{what makes a good measure of quantum correlations}? This question is typically answered by imposing a number of requirements that any such good measure should obey. Let us consider a one-sided measure $Q^{A}(\rho^{AB})$, defined by a real non-negative function acting on quantum states $\rho^{AB}$. One natural requirement is that
\begin{itemize}
\item{$Q^{A}(\chi^{AB}_{\text{cq}}) = 0$,}
\end{itemize}
i.e., that our measure is zero for classically correlated states. We should also expect that quantum correlations are not dependent upon the local bases of $A$ and $B$, which manifests as invariance under local unitaries $U^{A}$ on $A$ and $U^{B}$ on $B$,
\begin{itemize}
\item{$Q^{A}(\rho^{AB}) = Q^{A}([U^{A}\otimes U^{B} ]\rho^{AB}[U^{A}\otimes U^{B} ]^{\dagger})$.}
\end{itemize}
As we have already pointed out, entanglement and quantum correlations become the same phenomenon for pure states $\ket{\psi^{AB}}$, hence it is sensible to require that a measure of quantum correlations should reduce to a measure of entanglement for pure states,
\begin{itemize}
\item{$Q^{A}(\ket{\psi^{AB}}) = E^{A:B}(\ket{\psi^{AB}})$ for some entanglement measure  $E^{A:B}(\rho^{AB})$.}
\end{itemize}
Similar desiderata can be imposed for two-sided measures of quantum correlations $Q^{AB}(\rho^{AB})$.
However, so far we have not specified how our measure of quantum correlations should behave under dynamics of the system. In the case of entanglement, it is typically required that a measure should not increase under local operations and classical communication (LOCC) $\mathcal{E}^{AB}_{\text{LOCC}}$, i.e., $E^{A:B}(\mathcal{E}^{AB}_{\text{LOCC}}[\rho^{AB}]) \leq E^{A:B}(\rho^{AB})$~\cite{vedral1997quantifying,horodecki2009quantum}. In other words, one should not be able to generate entanglement by LOCC, the archetypal operations that spatially separated laboratories are limited to. This requirement is typically called monotonicity, and finding a comparable one for quantum correlations is tricky. For one-sided measures, it can be required that any local operation on subsystem $B$ should not be able to increase the quantum correlations from the perspective of subsystem $A$ \cite{streltsov2011linking}, that is,
\begin{itemize}
\item{$Q^{A}(\mathbb{I}^{A} \otimes \mathcal{E}^{B} [\rho^{AB}]) \leq Q^{A}(\rho^{AB})$ for any local operation $\mathcal{E}^{B}$ on $B$.}
\end{itemize}
Unfortunately, this cannot be the only monotonicity requirement, since it only specifies the local operations on $B$. Identifying the most meaningful and complete set of operations under which a good measure of quantum correlations should be monotone is currently an open question. We point the reader to~\cite{adesso2016measures} for a deeper explanation.

\section{\!\!\!Outlook}

We are going to be relying increasingly on the quantum world as technologies evolve during the 21$^{\text{st}}$ century, so it is certainly worthwhile to develop a good understanding of the quantum-classical boundary. In this manuscript we have focused on the most general type of quantum correlations between spatially separated parties. Whilst a promising topic, it is still very much in its infancy, with a plethora of interesting and open questions yet to be answered. From the theoretical side, perhaps the most pressing question is to identify a physically motivated set of ``free operations'' under which to impose monotonicity for measures of quantum correlations. This can be achieved by treating quantum correlations as a {\it resource}, within the framework of resource theories \cite{horodecki2013quantumness}. Experimentally, we have yet to witness  compelling evidence for the practical role of quantum correlations beyond entanglement in relevant quantum technologies, even though proof-of-principle demonstrations, e.g.~in the context of quantum metrology, are particularly promising \cite{girolami2014quantum}. In this respect, while the number of insightful operational interpretations for measures of quantum correlations has grown substantially in recent years \cite{adesso2016measures,georgescu2014apple}, {\em killer applications} are perhaps still waiting to be devised. 
It is hoped that by raising the awareness of these concepts within the wider quantum information community, we can begin to truly appreciate the foundational role and power of non-classical correlations beyond entanglement.

There are still many topics within the study of quantum correlations that we have  not had the opportunity to cover here. Foremost amongst which is the extensive research on their dynamics in open quantum systems, which shows that quantum correlations are generally more resilient than entanglement to the effects of typical sources of noise and decoherence~\cite{maziero2009classical,mazzola2010sudden}, a promising feature for any quantum technology. We have also neither discussed the role of quantum correlations in quantum computing~\cite{datta2008quantum,merali2011nature} and cryptography~\cite{pirandola2014quantum}, nor the quantification of quantum correlations among more than two parties~\cite{rulli2011global} or in continuous variable systems \cite{adesso2010quantum}. Nevertheless, there is a wealth of resources available to fill these gaps~\cite{modi2012classical,streltsov2015quantum,adesso2016measures,spehner2014quantum,roga2015geometric}. We hope to have passed on to the reader our enthusiasm for this young and blossoming field at the very core of quantum mechanics, and look forward to future progress.


\renewcommand{\baselinestretch}{0.95}


\begin{thebibliography}{42}
\expandafter\ifx\csname natexlab\endcsname\relax\def\natexlab#1{#1}\fi
\expandafter\ifx\csname bibnamefont\endcsname\relax
  \def\bibnamefont#1{#1}\fi
\expandafter\ifx\csname bibfnamefont\endcsname\relax
  \def\bibfnamefont#1{#1}\fi
\expandafter\ifx\csname citenamefont\endcsname\relax
  \def\citenamefont#1{#1}\fi
\expandafter\ifx\csname url\endcsname\relax
  \def\url#1{\texttt{#1}}\fi
\expandafter\ifx\csname urlprefix\endcsname\relax\def\urlprefix{URL }\fi
\providecommand{\bibinfo}[2]{#2}
\providecommand{\eprint}[2][]{\url{#2}}

{\small{
\bibitem{streltsov2016review}
\bibinfo{author}{\bibfnamefont{A.}~\bibnamefont{Streltsov}},
  \bibinfo{author}{\bibfnamefont{G.}~\bibnamefont{Adesso}}, \bibnamefont{and}
  \bibinfo{author}{\bibfnamefont{M.~B.} \bibnamefont{Plenio}},
  \bibinfo{journal}{arXiv preprint arXiv:1609.02439}  (\bibinfo{year}{2016}).

\bibitem{horodecki2009quantum}
\bibinfo{author}{\bibfnamefont{R.}~\bibnamefont{Horodecki}},
  \bibinfo{author}{\bibfnamefont{P.}~\bibnamefont{Horodecki}},
  \bibinfo{author}{\bibfnamefont{M.}~\bibnamefont{Horodecki}},
  \bibnamefont{and}
  \bibinfo{author}{\bibfnamefont{K.}~\bibnamefont{Horodecki}},
  \bibinfo{journal}{Rev. Mod. Phys.} \textbf{\bibinfo{volume}{81}},
  \bibinfo{pages}{865} (\bibinfo{year}{2009}).

\bibitem{modi2012classical}
\bibinfo{author}{\bibfnamefont{K.}~\bibnamefont{Modi}},
  \bibinfo{author}{\bibfnamefont{A.}~\bibnamefont{Brodutch}},
  \bibinfo{author}{\bibfnamefont{H.}~\bibnamefont{Cable}},
  \bibinfo{author}{\bibfnamefont{T.}~\bibnamefont{Paterek}}, \bibnamefont{and}
  \bibinfo{author}{\bibfnamefont{V.}~\bibnamefont{Vedral}},
  \bibinfo{journal}{Rev. Mod. Phys.} \textbf{\bibinfo{volume}{84}},
  \bibinfo{pages}{1655} (\bibinfo{year}{2012}).

\bibitem{streltsov2015quantum}
\bibinfo{author}{\bibfnamefont{A.}~\bibnamefont{{Streltsov}}},
  \emph{\bibinfo{title}{{Quantum Correlations Beyond Entanglement and their
  Role in Quantum Information Theory}}} (\bibinfo{publisher}{SpringerBriefs in
  Physics}, \bibinfo{year}{2015}), arXiv:1411.3208v1.

\bibitem{adesso2016measures}
\bibinfo{author}{\bibfnamefont{G.}~\bibnamefont{Adesso}},
  \bibinfo{author}{\bibfnamefont{T.~R.} \bibnamefont{Bromley}},
  \bibnamefont{and}
  \bibinfo{author}{\bibfnamefont{M.}~\bibnamefont{Cianciaruso}},
  \bibinfo{journal}{J. Phys. A: Math. Theor.} \textbf{\bibinfo{volume}{49}}, \bibinfo{pages}{473001} (\bibinfo{year}{2016}).
  

\bibitem{ollivier2001quantum}
\bibinfo{author}{\bibfnamefont{H.}~\bibnamefont{Ollivier}} \bibnamefont{and}
  \bibinfo{author}{\bibfnamefont{W.~H.} \bibnamefont{Zurek}},
  \bibinfo{journal}{Phys. Rev. Lett.} \textbf{\bibinfo{volume}{88}},
  \bibinfo{pages}{017901} (\bibinfo{year}{2001}).

\bibitem{henderson2001classical}
\bibinfo{author}{\bibfnamefont{L.}~\bibnamefont{Henderson}} \bibnamefont{and}
  \bibinfo{author}{\bibfnamefont{V.}~\bibnamefont{Vedral}},
  \bibinfo{journal}{J. Phys. A: Math. Gen.} \textbf{\bibinfo{volume}{34}},
  \bibinfo{pages}{6899} (\bibinfo{year}{2001}).

\bibitem{piani2008no}
\bibinfo{author}{\bibfnamefont{M.}~\bibnamefont{Piani}},
  \bibinfo{author}{\bibfnamefont{P.}~\bibnamefont{Horodecki}},
  \bibnamefont{and}
  \bibinfo{author}{\bibfnamefont{R.}~\bibnamefont{Horodecki}},
  \bibinfo{journal}{Phys. Rev. Lett.} \textbf{\bibinfo{volume}{100}},
  \bibinfo{pages}{090502} (\bibinfo{year}{2008}).

\bibitem{cavalcanti2016quantum}
\bibinfo{author}{\bibfnamefont{D.}~\bibnamefont{Cavalcanti}} \bibnamefont{and}
  \bibinfo{author}{\bibfnamefont{P.}~\bibnamefont{Skrzypczyk}},
  \bibinfo{journal}{arXiv preprint arXiv:1604.00501 to appear in Rep. Prog.
  Phys.}  (\bibinfo{year}{2016}).

\bibitem{brunner2014bell}
\bibinfo{author}{\bibfnamefont{N.}~\bibnamefont{Brunner}},
  \bibinfo{author}{\bibfnamefont{D.}~\bibnamefont{Cavalcanti}},
  \bibinfo{author}{\bibfnamefont{S.}~\bibnamefont{Pironio}},
  \bibinfo{author}{\bibfnamefont{V.}~\bibnamefont{Scarani}}, \bibnamefont{and}
  \bibinfo{author}{\bibfnamefont{S.}~\bibnamefont{Wehner}},
  \bibinfo{journal}{Rev. Mod. Phys.} \textbf{\bibinfo{volume}{86}},
  \bibinfo{pages}{419} (\bibinfo{year}{2014}).

\bibitem{schrodinger1935discussion}
\bibinfo{author}{\bibfnamefont{E.}~\bibnamefont{Schr{\"o}dinger}}, in
  \emph{\bibinfo{booktitle}{Mathematical Proceedings of the Cambridge
  Philosophical Society}} (\bibinfo{organization}{Cambridge Univ Press},
  \bibinfo{year}{1935}), vol.~\bibinfo{volume}{31}, pp.
  \bibinfo{pages}{555--563}.

\bibitem{ferraro2010almost}
\bibinfo{author}{\bibfnamefont{A.}~\bibnamefont{Ferraro}},
  \bibinfo{author}{\bibfnamefont{L.}~\bibnamefont{Aolita}},
  \bibinfo{author}{\bibfnamefont{D.}~\bibnamefont{Cavalcanti}},
  \bibinfo{author}{\bibfnamefont{F.}~\bibnamefont{Cucchietti}},
  \bibnamefont{and} \bibinfo{author}{\bibfnamefont{A.}~\bibnamefont{Ac\'in}},
  \bibinfo{journal}{Phys. Rev. A} \textbf{\bibinfo{volume}{81}},
  \bibinfo{pages}{052318} (\bibinfo{year}{2010}).

\bibitem{wu2009correlations}
\bibinfo{author}{\bibfnamefont{S.}~\bibnamefont{Wu}},
  \bibinfo{author}{\bibfnamefont{U.~V.} \bibnamefont{Poulsen}},
  \bibinfo{author}{\bibfnamefont{K.}~\bibnamefont{M{\o}lmer}},
  \bibnamefont{et~al.}, \bibinfo{journal}{Phys. Rev. A}
  \textbf{\bibinfo{volume}{80}}, \bibinfo{pages}{032319}
  (\bibinfo{year}{2009}).

\bibitem{divincenzo2004locking}
\bibinfo{author}{\bibfnamefont{D.~P.} \bibnamefont{DiVincenzo}},
  \bibinfo{author}{\bibfnamefont{M.}~\bibnamefont{Horodecki}},
  \bibinfo{author}{\bibfnamefont{D.~W.} \bibnamefont{Leung}},
  \bibinfo{author}{\bibfnamefont{J.~A.} \bibnamefont{Smolin}},
  \bibnamefont{and} \bibinfo{author}{\bibfnamefont{B.~M.}
  \bibnamefont{Terhal}}, \bibinfo{journal}{Phys. Rev. Lett.}
  \textbf{\bibinfo{volume}{92}}, \bibinfo{pages}{067902}
  (\bibinfo{year}{2004}).

\bibitem{lang2011entropic}
\bibinfo{author}{\bibfnamefont{M.~D.} \bibnamefont{Lang}},
  \bibinfo{author}{\bibfnamefont{C.~M.} \bibnamefont{Caves}}, \bibnamefont{and}
  \bibinfo{author}{\bibfnamefont{A.}~\bibnamefont{Shaji}},
  \bibinfo{journal}{Int. J. Quant. Inf.} \textbf{\bibinfo{volume}{9}},
  \bibinfo{pages}{1553} (\bibinfo{year}{2011}).

\bibitem{luo2010quantum}
\bibinfo{author}{\bibfnamefont{S.}~\bibnamefont{Luo}}, \bibinfo{journal}{Lett.
  Math. Phys.} \textbf{\bibinfo{volume}{92}}, \bibinfo{pages}{143}
  (\bibinfo{year}{2010}).

\bibitem{barnum1996noncommuting}
\bibinfo{author}{\bibfnamefont{H.}~\bibnamefont{Barnum}},
  \bibinfo{author}{\bibfnamefont{C.~M.} \bibnamefont{Caves}},
  \bibinfo{author}{\bibfnamefont{C.~A.} \bibnamefont{Fuchs}},
  \bibinfo{author}{\bibfnamefont{R.}~\bibnamefont{Jozsa}}, \bibnamefont{and}
  \bibinfo{author}{\bibfnamefont{B.}~\bibnamefont{Schumacher}},
  \bibinfo{journal}{Phys. Rev. Lett.} \textbf{\bibinfo{volume}{76}},
  \bibinfo{pages}{2818} (\bibinfo{year}{1996}).

\bibitem{wootters1982single}
\bibinfo{author}{\bibfnamefont{W.~K.} \bibnamefont{Wootters}} \bibnamefont{and}
  \bibinfo{author}{\bibfnamefont{W.~H.} \bibnamefont{Zurek}},
  \bibinfo{journal}{Nature} \textbf{\bibinfo{volume}{299}},
  \bibinfo{pages}{802} (\bibinfo{year}{1982}).

\bibitem{dieks1982communication}
\bibinfo{author}{\bibfnamefont{D.}~\bibnamefont{Dieks}},
  \bibinfo{journal}{Phys. Lett. A} \textbf{\bibinfo{volume}{92}},
  \bibinfo{pages}{271} (\bibinfo{year}{1982}).

\bibitem{brandao2015generic}
\bibinfo{author}{\bibfnamefont{F.~G. S.~L.} \bibnamefont{Brand{\~a}o}},
  \bibinfo{author}{\bibfnamefont{M.}~\bibnamefont{Piani}}, \bibnamefont{and}
  \bibinfo{author}{\bibfnamefont{P.}~\bibnamefont{Horodecki}},
  \bibinfo{journal}{Nature Commun.} \textbf{\bibinfo{volume}{6}},
  \bibinfo{pages}{7908} (\bibinfo{year}{2015}).

\bibitem{streltsov2013quantum}
\bibinfo{author}{\bibfnamefont{A.}~\bibnamefont{Streltsov}} \bibnamefont{and}
  \bibinfo{author}{\bibfnamefont{W.~H.} \bibnamefont{Zurek}},
  \bibinfo{journal}{Phys. Rev. Lett.} \textbf{\bibinfo{volume}{111}},
  \bibinfo{pages}{040401} (\bibinfo{year}{2013}).

\bibitem{streltsov2011linking}
\bibinfo{author}{\bibfnamefont{A.}~\bibnamefont{Streltsov}},
  \bibinfo{author}{\bibfnamefont{H.}~\bibnamefont{Kampermann}},
  \bibnamefont{and} \bibinfo{author}{\bibfnamefont{D.}~\bibnamefont{Bru{\ss}}},
  \bibinfo{journal}{Phys. Rev. Lett.} \textbf{\bibinfo{volume}{106}},
  \bibinfo{pages}{160401} (\bibinfo{year}{2011}).

\bibitem{piani2011all}
\bibinfo{author}{\bibfnamefont{M.}~\bibnamefont{Piani}},
  \bibinfo{author}{\bibfnamefont{S.}~\bibnamefont{Gharibian}},
  \bibinfo{author}{\bibfnamefont{G.}~\bibnamefont{Adesso}},
  \bibinfo{author}{\bibfnamefont{J.}~\bibnamefont{Calsamiglia}},
  \bibinfo{author}{\bibfnamefont{P.}~\bibnamefont{Horodecki}},
  \bibnamefont{and} \bibinfo{author}{\bibfnamefont{A.}~\bibnamefont{Winter}},
  \bibinfo{journal}{Phys. Rev. Lett.} \textbf{\bibinfo{volume}{106}},
  \bibinfo{pages}{220403} (\bibinfo{year}{2011}).

\bibitem{baumgratz2014quantifying}
\bibinfo{author}{\bibfnamefont{T.}~\bibnamefont{Baumgratz}},
  \bibinfo{author}{\bibfnamefont{M.}~\bibnamefont{Cramer}}, \bibnamefont{and}
  \bibinfo{author}{\bibfnamefont{M.~B.} \bibnamefont{Plenio}},
  \bibinfo{journal}{Phys. Rev. Lett.} \textbf{\bibinfo{volume}{113}},
  \bibinfo{pages}{140401} (\bibinfo{year}{2014}).

\bibitem{girolami2014observable}
\bibinfo{author}{\bibfnamefont{D.}~\bibnamefont{Girolami}},
  \bibinfo{journal}{Phys. Rev. Lett.} \textbf{\bibinfo{volume}{113}},
  \bibinfo{pages}{170401} (\bibinfo{year}{2014}).

\bibitem{marvian2014extending}
\bibinfo{author}{\bibfnamefont{I.}~\bibnamefont{Marvian}} \bibnamefont{and}
  \bibinfo{author}{\bibfnamefont{R.~W.} \bibnamefont{Spekkens}},
  \bibinfo{journal}{Nature Commun.} \textbf{\bibinfo{volume}{5}},
  \bibinfo{pages}{3821} (\bibinfo{year}{2014}).

\bibitem{braunstein1994statistical}
\bibinfo{author}{\bibfnamefont{S.~L.} \bibnamefont{Braunstein}}
  \bibnamefont{and} \bibinfo{author}{\bibfnamefont{C.~M.} \bibnamefont{Caves}},
  \bibinfo{journal}{Phys. Rev. Lett.} \textbf{\bibinfo{volume}{72}},
  \bibinfo{pages}{3439} (\bibinfo{year}{1994}).

\bibitem{girolami2014quantum}
\bibinfo{author}{\bibfnamefont{D.}~\bibnamefont{Girolami}},
  \bibinfo{author}{\bibfnamefont{A.~M.} \bibnamefont{Souza}},
  \bibinfo{author}{\bibfnamefont{V.}~\bibnamefont{Giovannetti}},
  \bibinfo{author}{\bibfnamefont{T.}~\bibnamefont{Tufarelli}},
  \bibinfo{author}{\bibfnamefont{J.~G.} \bibnamefont{Filgueiras}},
  \bibinfo{author}{\bibfnamefont{R.~S.} \bibnamefont{Sarthour}},
  \bibinfo{author}{\bibfnamefont{D.~O.} \bibnamefont{Soares-Pinto}},
  \bibinfo{author}{\bibfnamefont{I.~S.} \bibnamefont{Oliveira}},
  \bibnamefont{and} \bibinfo{author}{\bibfnamefont{G.}~\bibnamefont{Adesso}},
  \bibinfo{journal}{Phys. Rev. Lett.} \textbf{\bibinfo{volume}{112}},
  \bibinfo{pages}{210401} (\bibinfo{year}{2014}).

\bibitem{vonneumann1932mathematical}
\bibinfo{author}{\bibfnamefont{J.}~\bibnamefont{von Neumann}},
  \emph{\bibinfo{title}{Mathematical Foundations of Quantum Mechanics}}
  (\bibinfo{publisher}{Springer-Verlag}, \bibinfo{address}{Berlin},
  \bibinfo{year}{1932}).

\bibitem{adesso2014experimental}
\bibinfo{author}{\bibfnamefont{G.}~\bibnamefont{Adesso}},
  \bibinfo{author}{\bibfnamefont{V.}~\bibnamefont{D’Ambrosio}},
  \bibinfo{author}{\bibfnamefont{E.}~\bibnamefont{Nagali}},
  \bibinfo{author}{\bibfnamefont{M.}~\bibnamefont{Piani}}, \bibnamefont{and}
  \bibinfo{author}{\bibfnamefont{F.}~\bibnamefont{Sciarrino}},
  \bibinfo{journal}{Phys. Rev. Lett.} \textbf{\bibinfo{volume}{112}},
  \bibinfo{pages}{140501} (\bibinfo{year}{2014}).

\bibitem{vedral1997quantifying}
\bibinfo{author}{\bibfnamefont{V.}~\bibnamefont{Vedral}},
  \bibinfo{author}{\bibfnamefont{M.~B.} \bibnamefont{Plenio}},
  \bibinfo{author}{\bibfnamefont{M.~A.} \bibnamefont{Rippin}},
  \bibnamefont{and} \bibinfo{author}{\bibfnamefont{P.~L.}
  \bibnamefont{Knight}}, \bibinfo{journal}{Phys. Rev. Lett.}
  \textbf{\bibinfo{volume}{78}}, \bibinfo{pages}{2275} (\bibinfo{year}{1997}).

\bibitem{spehner2014quantum}
\bibinfo{author}{\bibfnamefont{D.}~\bibnamefont{Spehner}}, \bibinfo{journal}{J.
  Math. Phys.} \textbf{\bibinfo{volume}{55}}, \bibinfo{pages}{075211}
  (\bibinfo{year}{2014}).

\bibitem{roga2015geometric}
\bibinfo{author}{\bibfnamefont{W.}~\bibnamefont{Roga}},
  \bibinfo{author}{\bibfnamefont{D.}~\bibnamefont{Spehner}}, \bibnamefont{and}
  \bibinfo{author}{\bibfnamefont{F.}~\bibnamefont{Illuminati}},
  \bibinfo{journal}{J. Phys. A: Math. Theor.} \textbf{\bibinfo{volume}{49}},
  \bibinfo{pages}{235301} (\bibinfo{year}{2016}).

\bibitem{horodecki2013quantumness}
\bibinfo{author}{\bibfnamefont{M.}~\bibnamefont{Horodecki}} \bibnamefont{and}
  \bibinfo{author}{\bibfnamefont{J.}~\bibnamefont{Oppenheim}},
  \bibinfo{journal}{Int. J. Mod. Phys. B} \textbf{\bibinfo{volume}{27}},
  \bibinfo{pages}{1345019} (\bibinfo{year}{2013}).

\bibitem{georgescu2014apple}
\bibinfo{author}{\bibfnamefont{I.}~\bibnamefont{Georgescu}},
  \bibinfo{journal}{Nature Phys.} \textbf{\bibinfo{volume}{10}},
  \bibinfo{pages}{474} (\bibinfo{year}{2014}).

\bibitem{maziero2009classical}
\bibinfo{author}{\bibfnamefont{J.}~\bibnamefont{Maziero}},
  \bibinfo{author}{\bibfnamefont{L.~C.} \bibnamefont{C\'eleri}},
  \bibinfo{author}{\bibfnamefont{R.~M.} \bibnamefont{Serra}}, \bibnamefont{and}
  \bibinfo{author}{\bibfnamefont{V.}~\bibnamefont{Vedral}},
  \bibinfo{journal}{Phys. Rev. A} \textbf{\bibinfo{volume}{80}},
  \bibinfo{pages}{044102} (\bibinfo{year}{2009}).

\bibitem{mazzola2010sudden}
\bibinfo{author}{\bibfnamefont{L.}~\bibnamefont{Mazzola}},
  \bibinfo{author}{\bibfnamefont{J.}~\bibnamefont{Piilo}}, \bibnamefont{and}
  \bibinfo{author}{\bibfnamefont{S.}~\bibnamefont{Maniscalco}},
  \bibinfo{journal}{Phys. Rev. Lett.} \textbf{\bibinfo{volume}{104}},
  \bibinfo{pages}{200401} (\bibinfo{year}{2010}).

\bibitem{datta2008quantum}
\bibinfo{author}{\bibfnamefont{A.}~\bibnamefont{Datta}},
  \bibinfo{author}{\bibfnamefont{A.}~\bibnamefont{Shaji}}, \bibnamefont{and}
  \bibinfo{author}{\bibfnamefont{C.~M.} \bibnamefont{Caves}},
  \bibinfo{journal}{Phys. Rev. Lett.} \textbf{\bibinfo{volume}{100}},
  \bibinfo{pages}{050502} (\bibinfo{year}{2008}).

\bibitem{merali2011nature}
\bibinfo{author}{\bibfnamefont{Z.}~\bibnamefont{Merali}},
  \bibinfo{journal}{Nature} \textbf{\bibinfo{volume}{474}}, \bibinfo{pages}{24}
  (\bibinfo{year}{2011}).

\bibitem{pirandola2014quantum}
\bibinfo{author}{\bibfnamefont{S.}~\bibnamefont{Pirandola}},
  \bibinfo{journal}{Sci. Rep.} \textbf{\bibinfo{volume}{4}},
  \bibinfo{pages}{6956} (\bibinfo{year}{2014}).

\bibitem{rulli2011global}
\bibinfo{author}{\bibfnamefont{C.~C.} \bibnamefont{Rulli}} \bibnamefont{and}
  \bibinfo{author}{\bibfnamefont{M.~S.} \bibnamefont{Sarandy}},
  \bibinfo{journal}{Phys. Rev. A} \textbf{\bibinfo{volume}{84}},
  \bibinfo{pages}{042109} (\bibinfo{year}{2011}).

\bibitem{adesso2010quantum}
\bibinfo{author}{\bibfnamefont{G.}~\bibnamefont{Adesso}} \bibnamefont{and}
  \bibinfo{author}{\bibfnamefont{A.}~\bibnamefont{Datta}},
  \bibinfo{journal}{Phys. Rev. Lett.} \textbf{\bibinfo{volume}{105}},
  \bibinfo{pages}{030501} (\bibinfo{year}{2010}).

}}
\end{thebibliography}
\end{document}